\documentclass[twocolumn]{aastex62}

\graphicspath{{./}{figures/}}

\received{November 11, 2018}
\revised{March 4, 2019}
\accepted{March 5, 2019}

\submitjournal{ApJ}

\shorttitle{Red Sequence Scatter}
\shortauthors{Connor et al.}

\begin{document}

\title{On the Origin of the Scatter in the Red Sequence: An Analysis of Four CLASH Clusters}

\correspondingauthor{Thomas Connor}
\email{tconnor@carnegiescience.edu}

\author[0000-0002-7898-7664]{Thomas Connor}
\affil{The Observatories of the Carnegie Institution for Science, 813 Santa Barbara St., Pasadena, CA 91101, USA}

\author[0000-0003-4727-4327]{Daniel D. Kelson}
\affiliation{The Observatories of the Carnegie Institution for Science, 813 Santa Barbara St., Pasadena, CA 91101, USA}

\author[0000-0002-2808-0853]{Megan Donahue}
\affiliation{Department of Physics and Astronomy, Michigan State University, East Lansing, MI 48823, USA}

\author[0000-0002-2733-4559]{John Moustakas}
\affiliation{Department of Physics and Astronomy, Siena College, 515 Loudon Road, Loudonville, NY 12211, USA}

\begin{abstract}
In clusters of galaxies, the red sequence is believed to be a consequence of a correlation between stellar mass and chemical abundances, with more massive galaxies being more metal-rich and, as a consequence, redder. However, there is a color scatter around the red sequence that holds even with precision photometry, implying that the galaxy population is more complicated than as described by a mass-metallicity relation. We use precision photometry from the Cluster Lensing and Supernova survey with Hubble (CLASH) to investigate what drives this scatter. In four CLASH clusters at $z=0.355 \pm 0.007$, we find that the optical-IR galaxy colors confirm the previously known trend of metallicity along the red sequence but also show a strong connection between stellar age and red sequence offset, with ages ranging from 3 to 8 Gyr. Starting with fixed-age color-magnitude relations motivated by the mass-metallicity correlations of CLASH cluster galaxies, and by adjusting galaxy colors through stellar population models to put them all at the age of our red sequence, we are able to reduce the, e.g., $\textrm{F}625\textrm{W}-\textrm{F}814\textrm{W}$ scatter from 0.051 mag to 0.026 mag with median photometric errors of 0.029 mag. While we will extend this analysis to the full CLASH sample, in four clusters our technique already provides a color precision in near-total-light apertures to resolve the spread in stellar population formation ages that drives the scatter in the red sequence.
\end{abstract}


\section{Introduction} \label{sec:intro}

One of the defining properties of the population of galaxies in galaxy clusters is the red sequence -- a tight correlation between color and luminosity, with a slight trend for fainter galaxies to be bluer. This well-known relation \citep{1959PASP...71..106B}, alternatively known as the color-magnitude relation \citep{1977ApJ...216..214V}, has been seen in the stellar populations of early-type galaxies in groups and clusters throughout most of cosmic history. 

The red optical colors of galaxies can be explained equally well for one color through either enhanced metallicity or advanced age \citep[this is known as the age-metallicity degeneracy;][]{1994ApJS...95..107W}. While there was briefly some debate over which of these two factors drove the slope of the red sequence \citep[e.g.,][]{1996ASPC...98..287A}, \citet{1997A&A...320...41K} used the evolution of the observed red sequence at different redshifts to rule out age as the cause of the red sequence slope. In this context, the brighter red sequence galaxies are redder due to having more metals in their stellar atmospheres, implying a mass-dependency to how galaxies accumulate metals in their stellar populations \citep[e.g.,][]{1987A&A...173...23A}. 
\begin{figure*}
\includegraphics{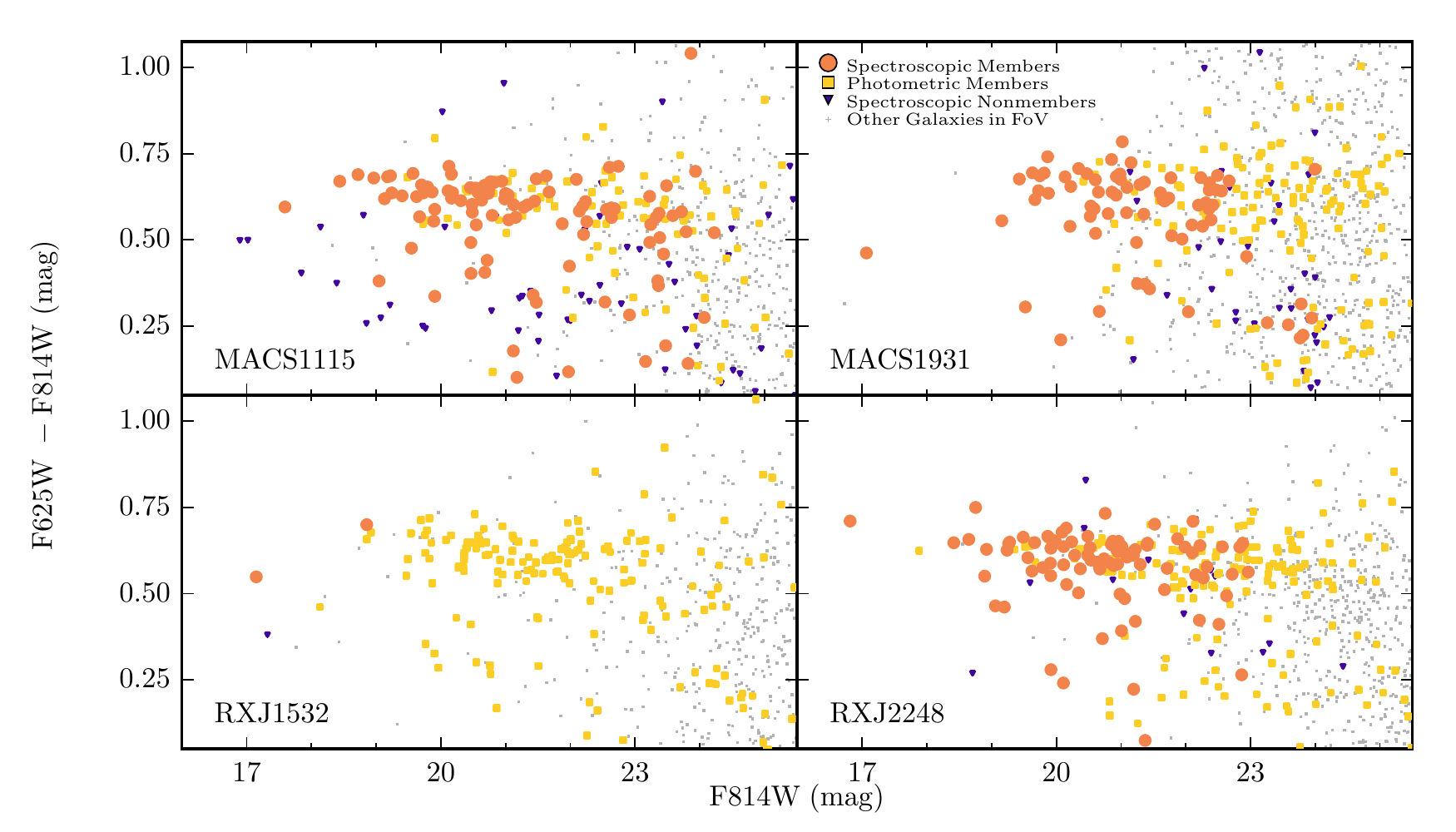}
\caption{Color-magnitude diagrams for the four clusters being studied in this work. Cluster members and interlopers are marked as indicated in the upper-right panel, where membership determinations and photometry are taken from \citet{2017ApJ...848...37C}. The spectroscopic coverage of RXJ 1532 is minimal. } \label{fig:four_cmds}
\end{figure*}

Since the work of \citet{1997A&A...320...41K}, questions about the red sequence have shifted toward how it forms. \citet{2007MNRAS.374..809D}, \citet{2008MNRAS.386.1045A}, \citet{2009ApJ...694.1349P}, and others have tried to constrain how the faint end of the red sequence is established. Searches for high-redshift clusters have led to observations of early red sequences \citep[e.g.,][]{2007MNRAS.377.1717K,2019A&A...622A.117S}, directly constraining the formation time more robustly than stellar population models of $z=0$ clusters. Additionally, there exists a growing interest in quantifying and understanding the intrinsic scatter of the red sequence \citep[e.g.,][]{2016arXiv161104671R,2018PASJ...70S..24N}. Understanding this scatter is the challenge we address in this work, although, as we will show, the scatter is related to the the formation timescales and mechanisms along the red sequence.

\citet{1992MNRAS.254..601B} showed that the color scatter was fairly small for the Coma and Virgo clusters; this analysis was extended to larger redshifts by, e.g., \citet{1997ApJ...483..582E} and \citet{1998ApJ...492..461S}. As optical colors show little change after a few Gyr have passed since the cessation of star formation \citep[][]{1976ApJ...203...52T}, this small scatter was taken as a sign that the cluster red sequence formed in a relatively small amount of time. 

Fundamentally, the question of scatter in the red sequence is a question of how well the red sequence describes the ensembles of stellar populations -- the larger the scatter, the less descriptive a red sequence model is for a specific galaxy. However, accurate measurement of scatter requires both a large number of galaxies and excellent photometry. Because of the former point, the less-populated bright end of the red sequence is much less useful than the middle and fainter ends, but, outside of the local universe, faint cluster populations are difficult to detect. 
Excellent photometry not only means precision -- photometric errors need to be smaller than any intrinsic scatter -- but accuracy; the intracluster light at 100 kpc has the color of a $10^{10}\ {\rm M}_\odot$ red sequence galaxy \citep{2018MNRAS.474.3009D}, so over- or under-subtracting that light will bias the galaxy photometry, particularly for faint galaxies.

\begin{figure*}
\includegraphics{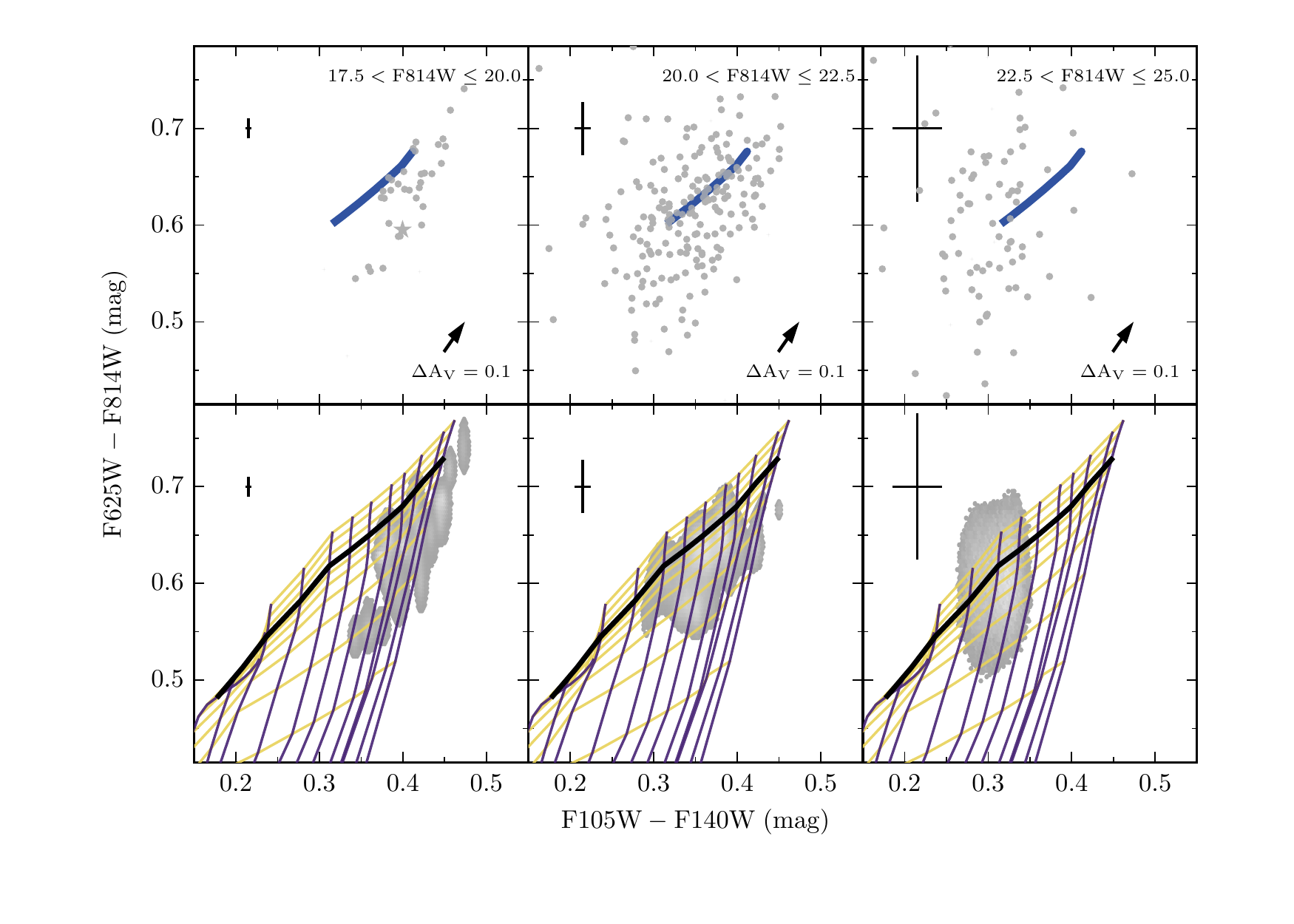}
\caption{Color-color diagram for red sequence galaxies in three brightness bins. From left to right, the bins cover $\left( 17.5, 20.0 \right]$, $\left( 20.0, 22.5 \right]$, and $\left( 22.5, 25.0 \right]$ in F814W magnitudes. In the bottom panels we show photometric density distributions by sampling each galaxy 100,000 times within its photometric uncertainties, then binning and plotting the total distributions. In the upper panels, the colors of our model red sequence are shown by the blue line. In the bottom panels, a grid of stellar population models are projected over the galaxy distribution; the black line traces an isochronal model with $z_f = 2.5$ (as used for our model red sequence), while purple and gold lines trace populations of constant metallicity and age, respectively. Here, higher metallicities and older populations are at the top right of the diagram. Representative errors for each bin are shown in the upper left of each panel.} \label{fig:cc_models}
\end{figure*}

To investigate the scatter in the red sequence, we utilize the photometric catalog of \citet{2017ApJ...848...37C}. This catalog provides excellent cluster galaxy photometry from the {\it Hubble Space Telescope} (HST) to magnitudes fainter than $M^* + 4.5$ for 25 clusters. As this photometry comes from the Cluster Lensing and Supernova survey with Hubble \citep[CLASH,][]{2012ApJS..199...25P}, there is photometry for each cluster covering the ultraviolet to the infrared. The broad photometric coverage is the third key ingredient to our analysis, as, in addition to a large sample of galaxies with precise photometry, the large color baseline enables us to break the age-metallicity degeneracy \citep[for previous uses of optical-IR colors to break the age-metallicity degeneracy, see, e.g.,][]{2001MNRAS.323..839S, 2005JKAS...38..145K, 2006MNRAS.367..339J}.

In this initial work we consider four clusters from CLASH that are all at redshift $z = 0.355 \pm 0.007$ (approximately 9.6 Gyr after the Big Bang). These clusters are MACS $1115.9{+}0129$ (hereafter MACS 1115), RXJ $1532.9{+}3021$ (hereafter RXJ 1532), MACS $1931.8{-}2635$ (hereafter MACS 1931), and RXJ $2248.7{-}4431$ (hereafter RXJ 2248). Because of their similar redshifts, they are seen when the age of the universe is the same to within ${\sim}120\ {\rm Myr}$. 
In this paper we quantify the properties of the red sequence for the combined sample and discuss how deviations from the red sequence are shaped by differing paths of galactic evolution. Throughout this work, we assume a flat $\Lambda$CDM cosmology with $\Omega_M = 0.3$ and ${\rm H}_0 = 70\ {\rm km}\ {\rm s}^{-1}\ {\rm Mpc}^{-1}$. To use a single population model, we make the simplifying approximation that all galaxies are at redshift $z = 0.35$. The scale at this redshift is $4.9\ \textrm{kpc}\ \textrm{arcsec}^{-1}$. All magnitudes in this work are AB.

\begin{figure*}
\includegraphics{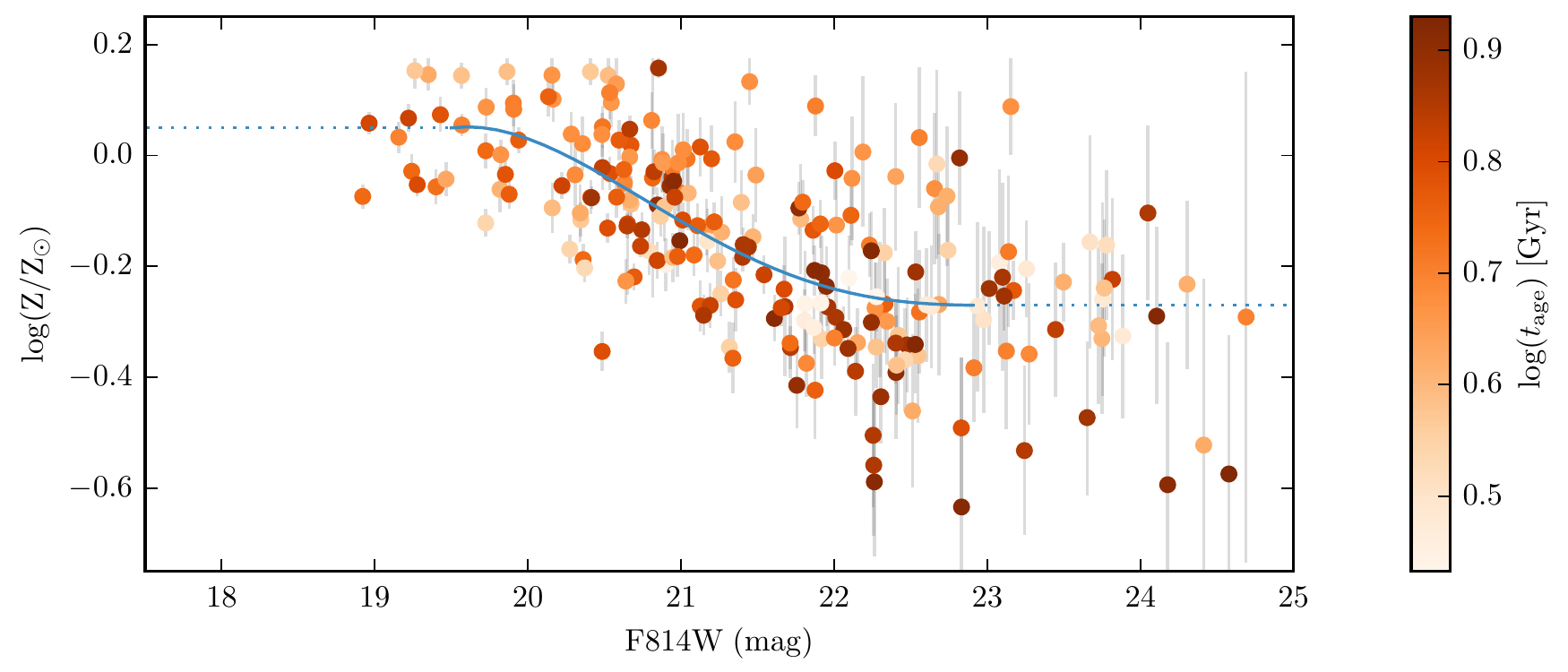}
\caption{Inferred galaxy metallicity as a function of galaxy brightness in the F814W filter. Galaxies are colored by the age of their stellar populations. A fit to these points, as described in the text, is shown by the blue line. The error bars show the $1\sigma$ distribution of metallicity values for each galaxy based on its photometric errors.} \label{fig:metal_mag}
\end{figure*}

\section{Data} \label{sec:data}

Galaxy photometry for the four CLASH clusters comes from the catalog of \citet{2017ApJ...848...37C}, generated using mode-based multi-scale filtering that disentangles the overlapping light profiles of galaxies, improving both the detection efficiency and photometric accuracy for cluster galaxies. All magnitudes in this catalog are effectively ``total light'' magnitudes (Kron-like), as described by the catalog paper. The images used by \citet{2017ApJ...848...37C} were first processed following \citet{2011ApJS..197...36K} and references therein. We use \citet{2017ApJ...848...37C}'s cluster membership assessments, which are based on spectroscopy when available and photometry when not. Redshifts for galaxies near the four clusters of interest come from \citet{2017ApJS..233...25A}, \citet{2012AJ....144...79G}, and \citet{2009A&A...499..357G}, as well as unpublished VLT-VIMOS redshifts (P. Rosati \& M. Nonino 2017, private communication) and a sample of unpublished redshifts from the IMACS-GISMO instrument on Magellan (D. D. Kelson 2017, private communication).

\begin{figure*}[p]
\includegraphics{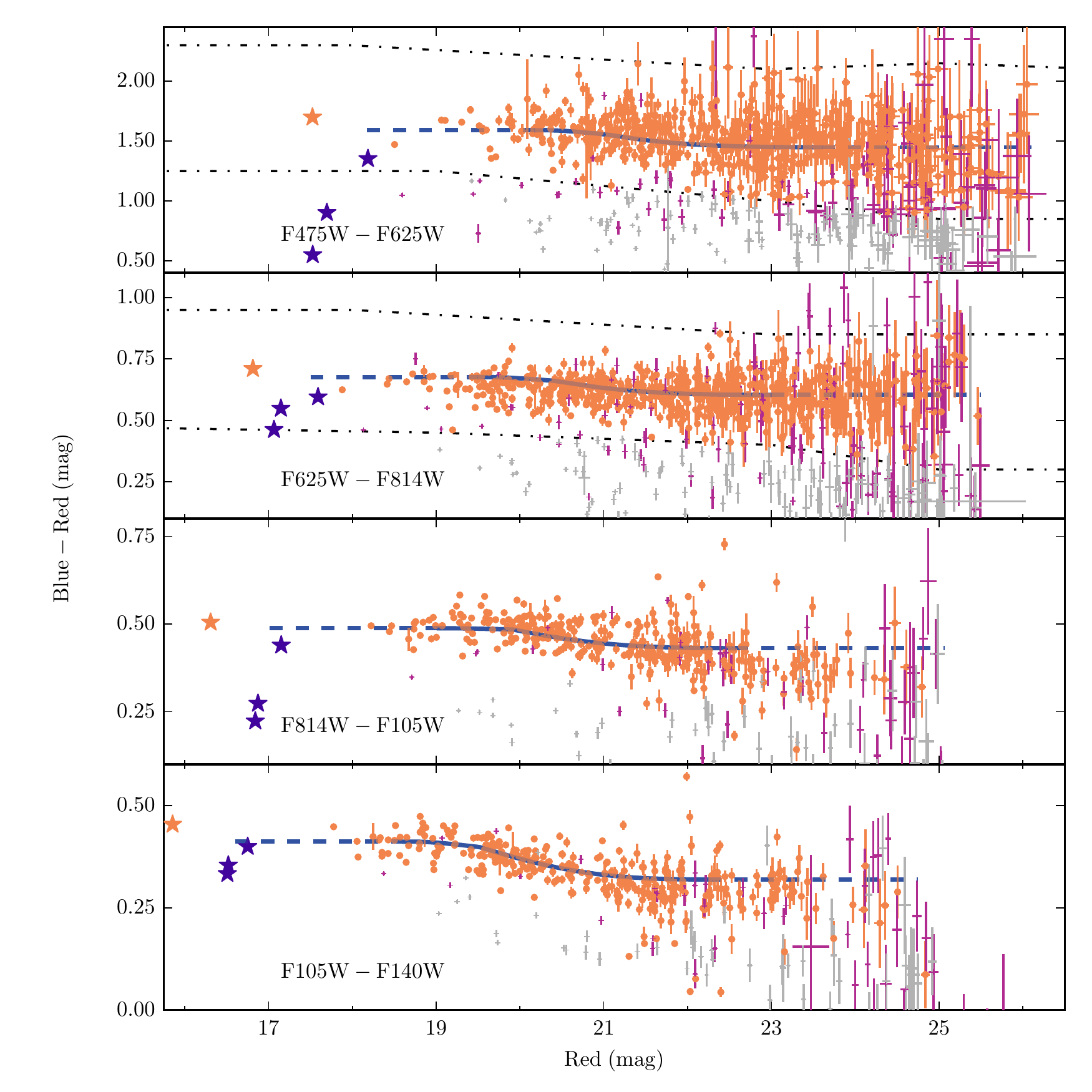}
\caption{Color-magnitude diagram of the combined four-cluster sample in four colors, as identified in the lower-left of each panel. Orange points are red sequence galaxies, gray are cluster members off the red sequence, and magenta are cluster members off the red sequence in one of the two color selection regions. These color selection regions are shown with dashed lines in the upper two panels. BCGs are marked by stars, and the three star-forming BCGs are colored dark blue. Our model red sequence is shown with the blue line; it is based on an assumption that the stellar populations of red sequence galaxies have the same age but that there is a decrease in the metallicity from the bright end to the faint end. The faint end cutoff is not due to photometric completeness but because \citet{2017ApJ...848...37C} imposed a hard cut at $\textrm{F}814\textrm{W}=25.5$\ mag.} \label{fig:all_four_fit}
\end{figure*}

We show ${\rm F814W} - {\rm F625W}$ color-magnitude diagrams (CMDs) for the four clusters in Figure \ref{fig:four_cmds}. Cluster ``members'' with and without spectroscopic redshifts are shown by orange circles and yellow squares, respectively. The red sequence is clearly visible in all four clusters, and it manifests in the same region of color-magnitude space. We note that the spectroscopic coverage for RXJ1532 is relatively non-existent, although the distribution of photometrically selected cluster members is in agreement with those of the clusters with more known redshifts. 

Our photometric catalogs include the brightest cluster galaxy (BCG) in each cluster. While these are all spectroscopically confirmed, they do not appear to be part of the red sequence; analyses by \citet{2015ApJ...805..177D} and \citet{2015ApJ...813..117F} found significant levels of ongoing star formation in the BCGs of MACS 1115, MACS 1931, and RXJ 1532. The BCG of RXJ 1532, in particular, has bright UV knots seen in the CLASH imaging \citep{2017ApJ...835..216D}. We therefore exclude these three BCGs from our analysis.

Combining optical and infrared colors helps break the age-metallicity degeneracy, so we begin our analysis of galaxy colors using $\textrm{F}475\textrm{W}-\textrm{F}625\textrm{W}$, $\textrm{F}625\textrm{W}-\textrm{F}814\textrm{W}$, $\textrm{F}814\textrm{W}-\textrm{F}105\textrm{W}$, and $\textrm{F}105\textrm{W}-\textrm{F}140\textrm{W}$. These five filters have minimal overlap and were imaged with the Advanced Camera for Surveys (ACS, \citealt{1998SPIE.3356..234F}: F475W, F625W, and F814W) and the Wide-Field Camera 3 (WFC3, \citealt{2008SPIE.7010E..1EK}: F105W and F140W). The ACS mosaicked field of view extends beyond a radius of ${\sim} 100\arcsec$ (${\sim}500$ kpc), while the WFC3 mosaicked field of view fully covers a circle of radius ${\sim} 60\arcsec$ (${\sim}300$ kpc). The use of all four filters therefore restricts us to only ${\sim}40\%$ that of the CLASH ACS field of view.

\section{Scatter Around the Red Sequence}
Traditionally, the red sequence has been modeled as a linear fit to a set of photometric points \citep[e.g.,][]{1998ApJ...501..571G} and, when multiple colors are being considered, they are fit independently \citep[e.g.,][]{2009MNRAS.394.2098S}. While this parameterization is useful for some contexts, it does not capture the underlying physical drivers of the relation, so it is inappropriate for evaluating deviations from the red sequence. We instead proffer the following: if the red sequence is caused by brighter galaxies being more metal rich, then the red sequence can be constructed by evaluating colors for model stellar populations with metallicities that decrease appropriately with decreasing stellar mass. By tying the normalization to a reasonable mass-metallicity relationship, we can start with a realistic ansatz red sequence that more appropriately captures any color changes.

Sadly, measurements of mass-metallicity relationships for cluster environments are relatively scarce \citep[often using nebular metallicities, which intrinsically miss red sequence galaxies, e.g.,][]{2019ApJ...872..192S}, so we instead use our own data to model an expectation for galaxy metallicities. While a single color is degenerate to mass and metallicity, two colors -- particularly an optical and an infrared color -- can break this degeneracy. We show in Figure \ref{fig:cc_models} our data in an IR-optical color-color plot; in the bottom panel, we overlay model predictions (described below) for various metallicity and stellar formation epoch ($z_f$) values. As changes in metallicity and age are mostly orthogonal in this diagram, we can use a galaxy's position to infer both parameters. This limits us to only those galaxies in the WFC3 field.

Model colors were generated using {\tt EzGal} \citep{2012PASP..124..606M}, assuming a \citet{2003PASP..115..763C} initial mass function and the stellar population synthesis models of \citet{2009ApJ...699..486C} and \citet{2010ApJ...712..833C}. We assume exponentially decaying star formation histories with decay time $\tau=1.0\ \textrm{Gyr}$. While only a limited subset of our model evaluations are shown in Figure \ref{fig:cc_models}, our interpolation grid used for metallicity estimation  is $101\times101$ and linearly spaced in the domain $z_f \in [0.8,5.8]$, $Z \in [0.0030,0.0300]$, where the solar metallicity is $Z_\odot=0.0200$. A fundamental assumption of this work is that the galaxies can be described by simple stellar populations (SSPs). While \citet{2011ApJ...726..110Z} found that some red sequence galaxies harbor ongoing star-formation, their targets were not in clusters. Similarly, \citet{2005A&A...443..435W} showed that star-forming red galaxies were in cluster outskirts, which we do not sample. Thus, we do not expect any significant contamination from non-SSP galaxies.

\begin{figure*}[ht]
\includegraphics{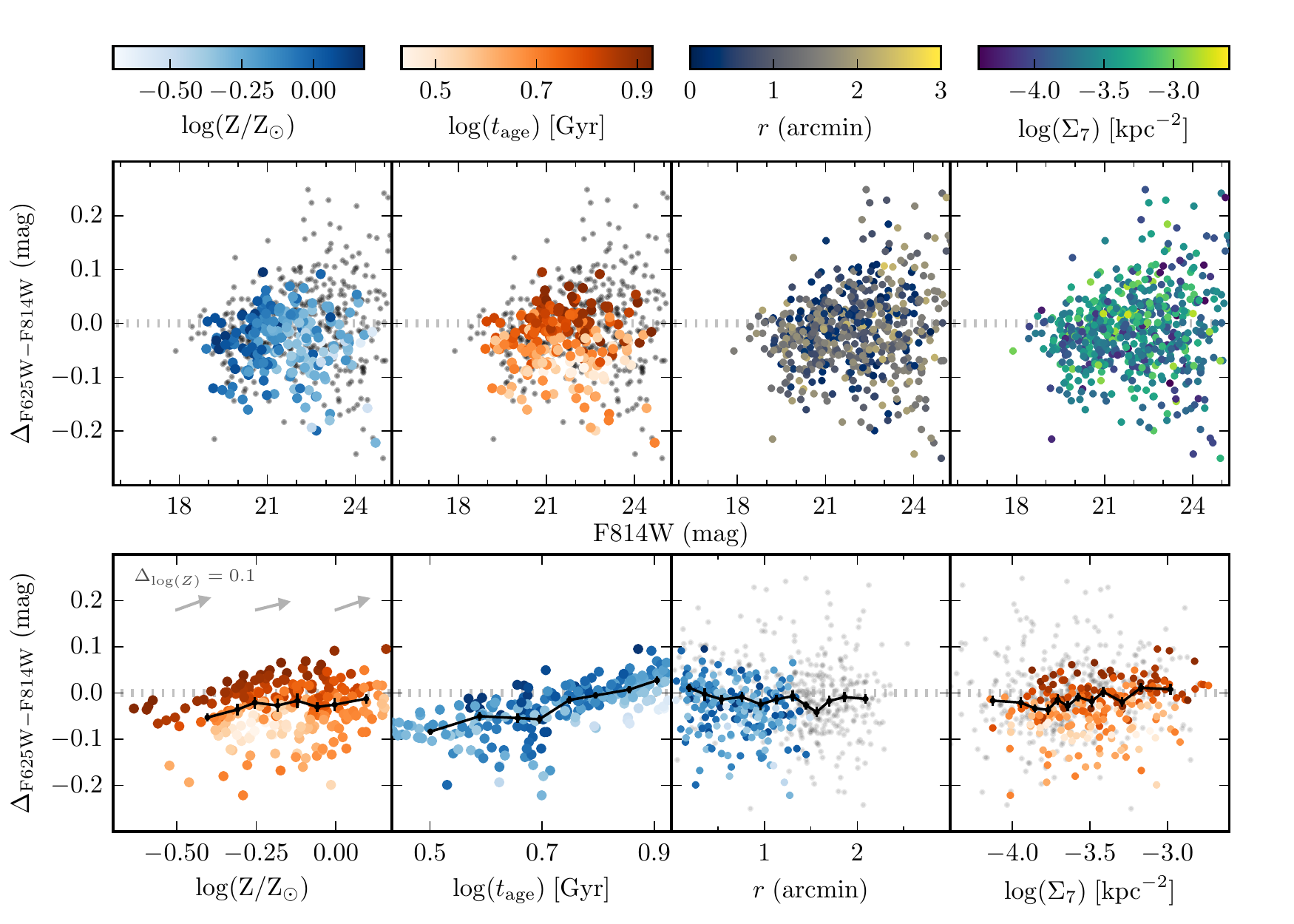}
\caption{Potential causes of offset from the red sequence. In the top four panels, we show the $\textrm{F625W}-\textrm{F814W}$ red sequence offsets vs F814W magnitudes. From left to right, galaxies are colored by their metallicity, stellar age, clustercentric radius, and local galaxy surface density. In the bottom panels, we show the red sequence offsets plotted against those same four parameters, with the same color convention as in the top panels. We equally divided the populations into 8 (metallicity and age) or 12 (distance and density) bins; the median trends in those bins, as well as standard errors in the medians, are shown with the black lines. There is no significant correlation with clustercentric radius and surface density, and metallicity variations are primarily along the red sequence. Stellar formation epoch, however, is a significant source of red sequence scatter. In the bottom left panel we demonstrate how metallicity variations at fixed age will impact these results.} \label{fig:offsets}
\end{figure*}

With estimates of galaxy metallicities, we compute a rough $\textrm{F}814\textrm{W}$ magnitude - metallicity relation using a cubic spline. We fit the spline using all galaxies with metallicities, but we only define a magnitude-metallicity relation for F814W magnitudes between 19.5 and 23 (the inflection points of our spline). For galaxies outside this range, we assume a level metallicity distribution from the endpoints outward. This fit is shown in Figure \ref{fig:metal_mag}.

As can be seen in Figure \ref{fig:metal_mag}, the metallicities of galaxies vary around the fitted relation, but we need to further reduce the uncertainties on metallicity to make any meaningful remarks on the scatter (with the further caveat that our metallicity uncertainties do not include dust, as discussed below). Beyond $F814W \gtrsim 20.7$ mag, the uncertainty in metallicity is equal to or larger than the scatter about the metallicity-magnitude relation. Brighter than this cutoff, the scatter is ${\sim}0.13\ \textrm{dex}$ in metallicity at fixed brightness.  

Nevertheless, it has been suggested \citep[e.g.,][]{1995ASPC...86..203W} that there is relation between observed metallicities and ages, such that more metal-rich populations will appear younger, which we do not see. Statistically, the correlation coefficient $r$ between age and metallicity is $r=-0.160^{+0.161}_{-0.075}$ for $t$ and $Z$ and $r=-0.130^{+0.165}_{-0.086}$ for $\log(t)$ and $\log(Z)$ (90\% confidence intervals). Our technique is able to avoid this issue for two primary reasons: we include IR information \citep[unlike, e.g.,][]{1994ApJS...95..107W} and we do not use the same filter twice in our color-color diagram, 
which prevents a single photometric error from shifting a galaxy's position in Figure \ref{fig:cc_models} significantly in both age and metallicity.

From this brightness-metallicity relation, we can determine expected colors in our CMDs for red sequence galaxies. The only remaining constraint in our models is the epoch of stellar formation. Here, we assume $z_f = 2.5$; while this value provides a good match to our data, it is not a fitted quantity. An important point is that while recent works \citep[e.g.][]{2019ApJ...871...83S} have found evidence for $z_f \geq 5$ in clusters, those value are for the first-forming galaxies, while we adopt a later $z_f$ to be representative of the entire red sequence population. 

The resulting model red sequences are shown with the four color photometry in Figure \ref{fig:all_four_fit}. As can be seen in Figure \ref{fig:all_four_fit}, our ansatz captures the topology of the red sequence in all four bands in a manner a linear fit would not.
 We also show exclusion regions, delimited by dot-dashed black lines. 
 
To further our goal of understanding red sequence scatter, we exclude those galaxies bluer or redder than the shown region in either color. These regions were defined by eye to exclude blue cloud galaxies. While this exclusion could limit our observed age range, the practical effect is that only five excluded galaxies have IR colors that would place them on our interpolation grid, and these five are all younger than $\log (t_{age}) \leq 0.73$.

We now turn to color residuals from the red sequence; here, we focus on the optical color $\textrm{F625W}-\textrm{F814W}$.
For each galaxy, we compute a metallicity and $z_f$ using the interpolation grid shown in Figure \ref{fig:cc_models}. Those galaxies without WFC3 coverage, or with colors outside of our grid, are skipped. We also compute a clustercentric radius and a local surface density, $\Sigma_7$, defined as 
\begin{equation}
\Sigma_7 = 8 / \left(\pi r_7^2 \right),
\end{equation}
where $r_7$ is the distance to the seventh-nearest neighbor among cluster members, in kpc. 

Deviations from the $\textrm{F625W}-\textrm{F814W}$ red sequence are shown in Figure \ref{fig:offsets}; in the top panel, deviations are shown against magnitude, and in the bottom panel against four parameters: metallicity, age, clustercentric radius, and local projected galaxy density.
No significant correlation is seen with clustercentric radius or projected galaxy density, implying that the current environment is not a first-order effect (aside from all of these galaxies being in clusters). 
Color residuals are markedly correlated with stellar population age, however, with the color offsets corresponding to a population of galaxies ${\sim}3$ to ${\sim}8$ Gyr old. There is also a minor trend of increasing color with increasing metallicity; a linear fit to red sequence color offset vs. log metallicity has a slope of $0.0571 \pm 0.0589\ {\rm mag}\ {\rm dex}^{-1}$ (99.87\% confidence intervals, generated from bootstrapping). In comparison, a fit to color offset vs. log age has a slope of $0.271 \pm 0.059\ {\rm mag}\ {\rm dex}^{-1}$ (99.87\% confidence intervals). Thus, while there appears to be a secondary trend with metallicity, it is not statistically significant at a $3\sigma$ level. 

Finally, we consider how much of the scatter can be removed by adjusting galaxies to the same age. Limiting ourselves to only those galaxies in the interpolation grid, the $\textrm{F}625\textrm{W}-\textrm{F}814\textrm{W}$ scatter is 0.051 mag, where scatter is $1.4826\times\textrm{MAD}$, the median absolute deviation. For each of these galaxies, we calculate a new color, assuming the same metallicity as measured but with a stellar formation redshift of $z_f=2.5$. These colors show a reduced scatter of 0.026 mag, comparable to the median uncertainty in color of 0.029 mag. Thus, by correcting for the range of formation epochs, we have reduced the scatter to within the photometric uncertainty.

\section{Considerations of Our Technique}
\begin{figure}[t!]
\includegraphics{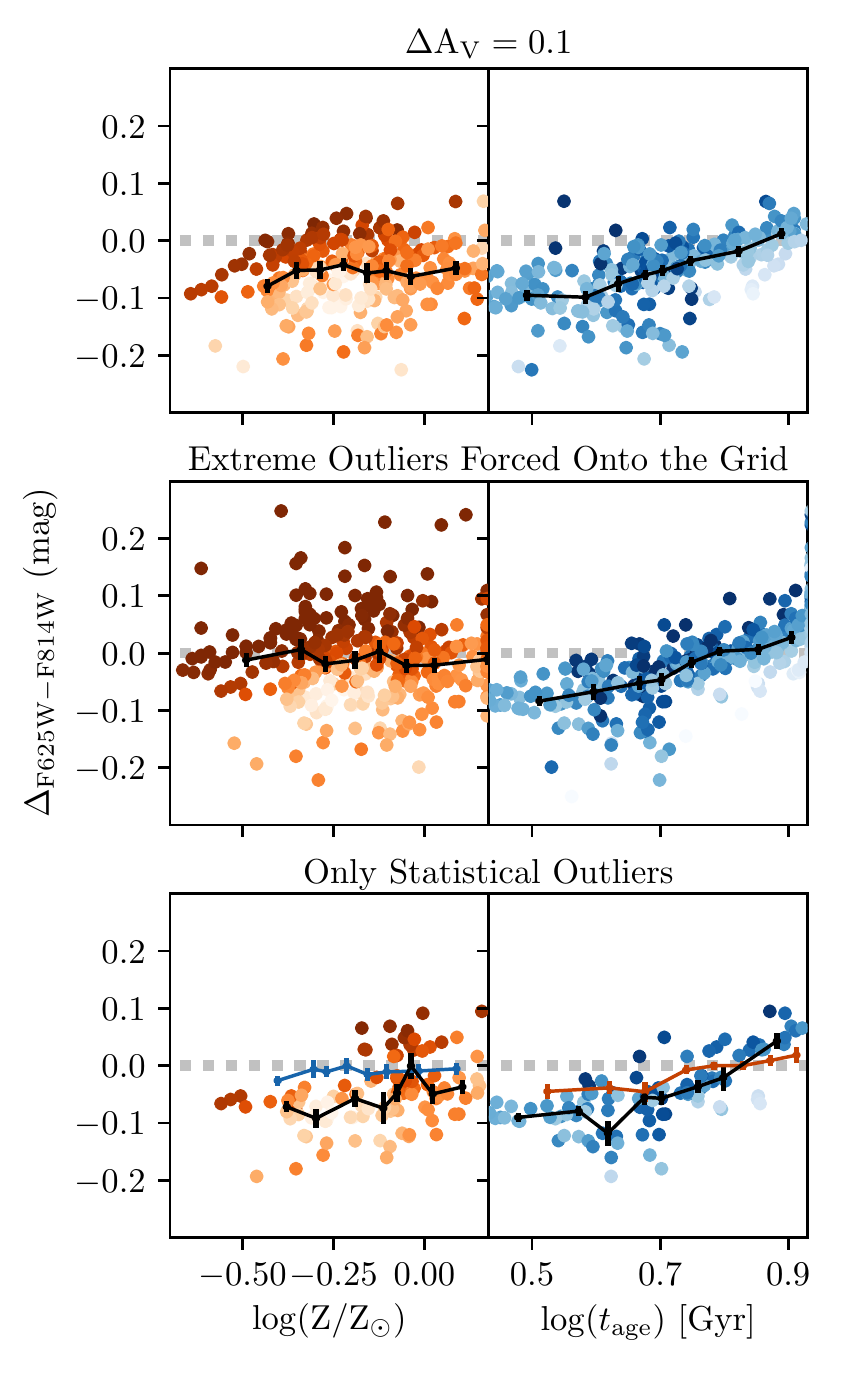}
\caption{The effects of several parameters on our results. Shown are the deviations from the red sequence with respect to metallicity and age (as is shown in the bottom left of Figure \ref{fig:offsets}). In the top, we show what the trend would look like if all galaxies were de-reddened, as described in the text. In the middle, we include all galaxies in the WFC3 field of view, and those outside the interpolation grid are given the values of the Voronoi cell they occupy. 
The bottom panel shows the distribution if all galaxies with photometric errors consistent with the red sequence are removed; the trend of the removed galaxies are shown by the colored lines. The colorbars are the same as in Figure \ref{fig:offsets}.} \label{fig:limitations}
\end{figure}

As this work is not intended to be a comprehensive description of the formation and evolution of the cluster red sequence, but rather a pilot study of a new paradigm for analyzing the red sequence, there are several limitations to be considered before generalizing our results and several assumptions we made whose effects we now quantify.
Two limits that can be addressed without refining this technique are the single redshift point and the accuracy of our model red sequence. The former point can be obviated by using the full CLASH sample, while the second will be addressed by future work properly calibrating a mass-metallicity relation for red sequence cluster galaxies.

We have not attempted to account for dust, which remains a minor source of uncertainty. Galactic dust and contamination from nearby galaxies were accounted for in the photometry, so that the primary concern is dust in each galaxy. In an analysis of (not cluster specific) elliptical galaxies, \citet{2011AJ....142..118S} found a median $A_V \approx 0.1$. This contribution, shown in Figure \ref{fig:cc_models}, produces a shift on the interpolation grid we used to characterize our galaxies. To test the severity of this effect, we consider a galaxy with colors $\textrm{F}105\textrm{W}-\textrm{F}140\textrm{W}=0.35$, $\textrm{F}625\textrm{W}-\textrm{F}814\textrm{W}=0.60$ and reddened by $A_V=0.1$, assuming a \citet{2000ApJ...533..682C} extinction law but with $R_V=3.02$, which is more characteristic of elliptical galaxies \citep{2007A&A...461..103P}. The difference due to added dust is $\Delta_{\log(t)}\sim0.05$ and a ${\sim}25\%$ increase in metallicity. Both effects are smaller than our observed trends. While dust won't change the properties of an individual galaxy significantly, we also consider what effect dust would have on our entire population. Applying the same $A_V=0.1$ reddening correction to our entire sample, then performing the same analysis as before, our observed scatter is 0.051 mag, which reduces to 0.029 mag  (uncertainty is 0.028 mag). Additionally, we show in Figure \ref{fig:limitations} that the same trend of color offsets being driven by age and metallicity is still present when applying a universal dust correction. It is possible that the effects of dust are not universal, and that there is an age- or metallicity-dependence to the amount of dust; while our data cannot discern this, we note again that the effects of dust on a sub-population of galaxies are still expected to be smaller than the observed trends.

Along a similar vein, 97 of the 320 galaxies within the WFC3 field of view are outside of our interpolation grid -- most of which are redder in the infrared color -- as shown in Figure \ref{fig:cc_models}. While we have assumed that these represent a small enough fraction of the population that the broader trends we have observed will hold regardless of how these galaxies behave, we are nevertheless quantifying a deviation while {\it a priori} ignoring the largest deviations. To see if this exclusion is affecting our results, we set the properties of all galaxies outside the interpolation grid to the closest point on the grid and repeat our analysis (in essence changing our model from an interpolation grid to a Voronoi diagram). Again, as shown in Figure \ref{fig:limitations}, the same age-dependent color-offset trend is observed, and adjusting galaxies to the same age as before reduces the observed scatter from 0.060 mag to 0.030 mag (median uncertainty is 0.033 mag). While the largest photometric deviations may be populating the tails of our age distribution, they are not driving the overall observational effect.

In this work, we have also limited ourselves to the Kron-like magnitudes from \citet{2017ApJ...848...37C}. Elliptical galaxies in clusters have been shown to have a color gradient \citep[e.g.,][]{1988A&A...203..217V,2000AJ....119.2134T}, with cores that are redder than their outskirts. This effect has even been seen in another cluster in the CLASH data set \citep{2018A&A...617A..34M}. As we are sampling galaxies covering three orders of magnitude in brightness, we can neither use apertures of fixed size (which do not scale) nor can we fit spatial profiles to each galaxy (despite using the same imaging data as we did, the analysis of \citealt{2018A&A...617A..34M} stopped ${\sim}3$ magnitudes brighter than in this work). 
Additionally, as noted by \citet{2001AJ....121.2413S}, measurements of the red sequence using fixed apertures are inherently biased due to the presence of color gradients. We therefore use the Kron-like magnitudes to both allow for dynamical scaling from large to small galaxies and to make our results less sensitive to the color gradient.
We also note that this work is focusing on color variations at a fixed magnitude; while the normalization and shape of the red sequence may vary by changing aperture sizes, the color differences of similar-sized galaxies still remains.

Our final consideration is the effect of photometric errors; while we have selected a data set that has precision photometry, the small uncertainties can nevertheless induce an apparent age spread. To see if our results are robust against this effect, we considered the age spread for galaxies with photometry at least $1\sigma$ from the red sequence in both  $\textrm{F}105\textrm{W}-\textrm{F}140\textrm{W}$ and $\textrm{F}625\textrm{W}-\textrm{F}814\textrm{W}$. This cut decreases our sample from 237 to 99 (in contrast to the ${\sim}25$ expected if galaxies were solely on the red sequence and scattered by random Gaussian errors). The distribution of off-sequence galaxies is shown in the bottom panel of Figure \ref{fig:limitations}, as is the behavior of those galaxies on the sequence. Even in this limited sample, the spread in colors along the age vector remains and the variation of colors at fixed metallicity is still correlated with inferred age. 
While there is still a small correlation between age and color offset for those galaxies with measured colors within $\pm 1 \sigma$ of the red sequence, this effect is much smaller than that observed for either the full sample or for just the off-sequence galaxies.

\section{Discussion}
We have shown that the red sequence at $z\approx0.35$ can be produced by the systemic variation of metallicity with stellar mass, with scatter coming from an age spread of ${\sim}4\,\textrm{Gyr}$, or ${\sim}40\%$ of a Hubble time. Such a large spread is consistent with a picture wherein cluster galaxies have stopped forming stars early, but a continual infall of galaxies from filaments is consistently reinforcing the red sequence population.  \citet{2015ApJ...812..138F} found a similar result, whereby the expected tightening of the red sequence through age was being balanced by the addition of new members. Likewise, \citet{2016arXiv161104671R} reported that the red sequence age scatter increases with time.

In Figure \ref{fig:offsets} we show an age scatter in the cluster population. The implication of this result is that the red sequence continues to gain galaxies with time. Consistent with this picture, some studies \citep[e.g.,][]{2007MNRAS.374..809D} have found an increasing fraction of faint galaxies with time \citep[although this is not without controversy; see e.g.,][]{2008MNRAS.386.1045A}. In a study of cluster luminosity functions, \citet{2009ApJ...700.1559R} reported that the bright end was passively evolving while the faint end is building up from $z=0.8$ to $z=0.4$.
\citet{2009A&A...499...47S} used line-strength indices and morphologies to argue for a 40\% growth in the faint red sequence from $z=0.75$ to $z=0.45$, independent of luminosity functions. In stacked spectra of quiescent galaxies, \citet{2014ApJ...792...95C} observed that the age evolution of the stacked lower-mass bin was not consistent with passive evolution, implying the addition of newly-quenched galaxies with time.

What we cannot do is directly compare scatter measurements with other studies, both for practical and philosophical reasons. Prior measures of the scatter have fitted a red sequence and reported the intrinsic scatter from that relation. This scatter is generally defined as $\sigma_{int}$, such that, when added in quadrature with the photometric errors, the goodness-of-fit becomes $\chi_\nu^2 = 1.0$ \citep[e.g.,][]{2004A&A...416..829L}. However, as we are not fitting a line but a model, we will inherently have different values of $\chi_\nu^2$, even for the same data. Also, as \citet{2016arXiv161104671R} discussed, this value is highly dependent on the photometric errors, and the intrinsic scatter can be reduced to 0 by simply taking worse observations. As defined, $\sigma_{int}$ also assumes a Gaussian distribution of colors around a nominal relation, but this assumption is not consistent with the scatter being an age spread around a nominal metallicity relation. 

Our results suggest that measuring the slope and scatter of red sequence galaxies is the wrong path to take for advancing our understanding of cluster galaxy evolution. The slope obfuscates the underlying mass-metallicity relation and the scatter is too easily distorted. Rather, we should be working toward two questions: ``What is the mass-metallicity relation of cluster galaxies?'' and ``What is the age distribution of red sequence members?'' These questions should, of course, also be considered in the context of halo masses, redshift, and galaxy size. Framing our investigations in this way not only directly connects to the underlying physics, but also allows comparisons across non-uniform data sets. 

\acknowledgments

{\scriptsize M.D. and T.C. were partially supported by STScI/NASA
awards HST-GO-13367 and  HST-GO-12065}

\facilities{HST}
\software{CosmoCalc \citep{2006PASP..118.1711W}, EzGal \citep{2012PASP..124..606M}}

\end{document}